\documentstyle[12pt]{article}

\newcommand{\sect}[1]{\setcounter{equation}{0}\section{#1}}

\newcommand{\EQ}{\begin{equation}}
\newcommand{\EN}{\end{equation}}
\newcommand{\bea}{\begin{eqnarray}}
\newcommand{\ena}{\end{eqnarray}}

\renewcommand{\a}{\alpha}
\renewcommand{\b}{\beta}

\renewcommand{\d}{\delta}

\newcommand{\pa}{\partial}
\newcommand{\g}{\gamma}
\newcommand{\G}{\Gamma}

\newcommand{\e}{\epsilon}

\newcommand{\z}{\zeta}

\newcommand{\k}{\kappa}

\renewcommand{\l}{\lambda}
\renewcommand{\L}{\Lambda}
\newcommand{\m}{\mu}

\newcommand{\n}{\nu}

\newcommand{\p}{\pi}

\newcommand{\r}{\rho}

\newcommand{\s}{\sigma}

\renewcommand{\t}{\tau}

\renewcommand{\o}{\omega}

\newcommand{\shalf}{\frac{1}{2}}

\begin{document}
 
\topmargin 0pt
\oddsidemargin 5mm
 
\renewcommand{\Im}{{\rm Im}\,}
\newcommand{\NP}[1]{Nucl.\ Phys.\ {\bf #1}}
\newcommand{\PL}[1]{Phys.\ Lett.\ {\bf #1}}
\newcommand{\NC}[1]{Nuovo Cimento {\bf #1}}
\newcommand{\CMP}[1]{Comm.\ Math.\ Phys.\ {\bf #1}}
\newcommand{\PR}[1]{Phys.\ Rev.\ {\bf #1}}
\newcommand{\PRL}[1]{Phys.\ Rev.\ Lett.\ {\bf #1}}
\newcommand{\MPL}[1]{Mod.\ Phys.\ Lett.\ {\bf #1}}
\renewcommand{\thefootnote}{\fnsymbol{footnote}}
\newpage
\begin{titlepage}
\vspace{2cm}
\begin{center}
{\bf{{\large RENORMALIZATION GROUP FLOWS IN SIGMA--MODELS COUPLED }}} \\
{\bf{{\large TO TWO--DIMENSIONAL DYNAMICAL GRAVITY }}} \\
\vspace{2cm}
{\large S. Penati} \footnote{E--mail address: penati@milano.infn.it},
{\large A. Santambrogio} \footnote{E--mail address: 
santambrogio@milano.infn.it} and 
{\large D. Zanon} \footnote{E--mail address: zanon@milano.infn.it} \\
\vspace{.2cm}
{\em Dipartimento di Fisica dell' Universit\`{a} di Milano and} \\
{\em INFN, Sezione di Milano, Via Celoria 16, I-20133 Milano, Italy}\\
\end{center}
\vspace{2cm}
\centerline{{\bf{Abstract}}}
\vspace{.5cm}
We consider a bosonic $\s$--model coupled to two--dimensional
gravity. In the semiclassical limit, $c\rightarrow -\infty$, we
compute the gravity dressing of the $\b$--functions at two--loop
order in the matter fields. We find that the corrections due to
the presence of dynamical gravity are {\em not} expressible 
simply in terms of a 
multiplicative factor as previously obtained at the one--loop level.
Our result indicates that the critical
points of the theory are nontrivially influenced and
modified by the induced gravity. 
\vfill
\noindent
IFUM--528--FT 
\hfill {May 1996}

\end{titlepage}
\renewcommand{\thefootnote}{\arabic{footnote}}
\setcounter{footnote}{0}
\newpage

\sect{Introduction}

A bosonic string theory can be described by free scalar fields coupled
to two--dimensional quantum gravity \cite{pol}. If the
dimensions of the system are critical ($d=26$), then the gravitational 
fields essentially decouple and act as a curved background.
If the system is not conformally invariant ($d\neq 26$),
gravity becomes dynamical and leads to important
modifications of the scaling properties of the matter which it
couples to \cite{polzam}, \cite{DK}.

Recently a number of papers have studied the 
gravitational dressing of the renormalization group equations
for two--dimensional models. 
In particular it has been shown
that the $\b$--function is multiplicatively 
renormalized at least at one--loop order in the matter fields
\cite{KKP},
\cite{million}, \cite{TKS}
\EQ
\b^{(1)}_G=\frac{\k+2}{\k+1}~ \b_0^{(1)}
\label{betaG}
\EN
where $\k$ is the central charge of the gravitational $SL(2R)$
Kac--Moody algebra
\EQ
\k+2= \frac{1}{12}[c-13-\sqrt{(1-c)(25-c)}]
\label{levelKM}
\EN
and $\b_0$ denotees the $\b$--function in the absence of gravity.
The above result has been checked in several examples and tested
using different quantization methods of gravity \cite{polzam},
\cite{DK}, \cite{KKP}, \cite{million}, \cite{TKS}. In all cases
the multiplicative renormalization factor appeared to be
one--loop universal.
Beyond lowest order of perturbation theory the one coupling case
has been examined \cite{dorn}: based on 
considerations in conformal gauge it has been argued that 
the gravitational
dressing of the nextleading order contribution to the
$\b$--function is not in the multiplicative form as in
eq. (\ref{betaG}). 

The motivation for the present work was to study the problem at the
two--loop level in a quantitative manner on a general 
setting. We have evaluated the
gravitational corrections to a bosonic $\s$--model 
$\b$--function at two loops in the matter fields.
The calculation has been performed following the
quantization procedure of refs. \cite{KKN}, \cite{TKS}. The
system is formulated in $n=2-2\e$ dimensions
with the gravitational conformal mode explicitly
separated. This provides a natural definition of 
{\em conformal} gauge in $n$ dimensions. The presence of 
the Liouville field modifies the physical scale with
respect to the standard renormalization scale $\m$. Indeed
the physical scale is defined through the only dimensionful
object in the theory, i.e. the gravitational cosmological
term
\EQ
\L_B \int d^nx \sqrt{-g}
\EN
where $\L_B$ is the bare cosmological constant. 
The renormalized constant defined by
\EQ
\L_B=\m^n\L_RZ
\EN
contains in general a divergent renormalization factor $Z$ .
Thus it is clear that, if renormalized
by quantum corrections,
the gravitational cosmological term acquires an anomalous dimension
and it modifies the relevant scale
to be used in the calculation of the dressed $\b$--function
\cite{KKP}, \cite{TKS}, \cite{GZ}
\EQ
\b_G=\frac{\pa ln\m}{\pa ln(\L_R^{-\frac{1}{n}})} \b
\label{beta}
\EN
being $\b$ the function which gives the change of the 
renormalized couplings with respect to the standard
mass scale $\m$.
In the presence of induced gravity one has
$\b=\b_0+\b_g$, having indicated with $\b_0$ the 
contribution to the beta function in the absence of gravity
and with $\b_g$ the corrections due to the gravity-matter
couplings. At one loop in the matter fields and to first order
in the semiclassical limit, $c\rightarrow-\infty$, it has been
shown that $\b^{(1)}_g=0$ so that $\b^{(1)}=\b_0^{(1)}$
\cite{TKS}.
Moreover the prefactor on the r.h.s. of eq. (\ref{beta})
has been computed  \cite{KKN} and complete agreement 
has been obtained at the one--loop order with the 
expectation in (\ref{betaG}).
We have extended the calculation at the two--loop level
in the matter fields: 
we have found that the multiplicative dressing is not
maintained, a rather compelling 
indication that the gravitational renormalization 
is not universal and that the fixed points of the
theory are nontrivially affected by the surrounding
dynamical gravity.

Our paper is organized as follows: in the next section
we describe the model and the approach we have followed
to perform the calculation. In section three we briefly
summarize the results at the one--loop level and illustrate
the general procedure for subtraction of infrared 
divergences. A careful treatement of this problem is
crucial for disentangling infrared versus ultraviolet 
divergences. 
The gravity dressed two--loop
matter contributions are presented in section four.
The final section contains some conclusive remarks,
while the two appendices give some details of the relevant
calculations.

\sect{The model and the general approach}

In order to obtain a quantum theory of gravitation in two dimensions, 
we start using a formulation of the theory in $n=2-2\e$
dimensions, following closely the work 
presented in ref. \cite{KKN}. There it has been shown
that within this approach, in the strong coupling regime and in the 
$\e\rightarrow 0$ limit, one reproduces the exact results of
two--dimensional quantum gravity \cite{polzam}, \cite{DK}.
We consider in $n$--dimension the
Hilbert--Einstein action given by
\EQ
S_{G}=\frac{1}{16\p G_0} \int d^nx \sqrt{-g}~R
\label{actionG}
\EN
where $G_0$ denotes the bare gravitational constant.
It can be reexpressed
in terms of a background metric, $\hat{g}_{\m\n}$, and
quantum fields, $h_{\m\n}$ and $\Phi$, defined in such a way that
\EQ
g_{\m\n}=\hat{g}_{\m\r} (e^h)^{\r}_{~\n} e^{-\frac{\Phi}{\e}}
\qquad \qquad \qquad h^{\m}_{~\m}=0
\EN
The quantum--gauge invariance under general coordinate 
transformations
is fixed introducing the following gauge--fixing term
\EQ
S_{GF}= -\frac{1}{32\p G_0} \int d^nx \sqrt{-\hat{g}}~
\left[\hat{D}^{\n} h_{\m\n}- \partial_{\m}\Phi\right]
\left[\hat{D}_{\r} h^{\m\r}- \partial^{\m}\Phi\right]
\label{actionGF}
\EN
In (\ref{actionGF}) $\hat{D}$ is the background covariant derivative
and indices are raised and lowered using the background metric.
The corresponding ghost action is given by
\bea
&&S_{ghosts}=\frac{1}{16\p G_0}\int d^nx\sqrt{-\hat{g}}~
\left[\psi^{*\m}\hat{D}^{\n}\hat{D}_{\n}\psi_{\m}
-\psi^{*\m}\hat{R}^{\n}_{~\m}\psi_{\n}+\right. \nonumber\\
&&~~~~~~~~~~~~~~~~~~~~~~~~\left.+\pa^{\n}\Phi\hat{D}_{\m}
\psi^{*\m}\psi_{\n}-\hat{D}_{\s} h^{\n}_{~\m} \hat{D}_{\n}
\psi^{*\m}\psi^{\s} +O(\e^2)\right]
\label{actionghosts}
\ena
Expanding the background around the flat metric,
from the quadratic terms in (\ref{actionG}) and (\ref{actionGF}), 
one easily obtains the propagators for the quantum fields
$h_{\m\n}$ and $\Phi$
\bea
&&\langle h_{\m\n} h_{\r\s} \rangle   = 
 16 \pi G_0 ~h_{\mu \nu  \r \sigma} 
~\frac{1}{q^2} \nonumber\\
&&\langle\Phi\Phi\rangle=16\p G_0\frac{2\e}{ n} \frac{1}{q^2}
\label{propagators}
\ena
where we have defined
\EQ
h_{\mu \nu  \r \sigma}\equiv\d_{\m\r}
\d_{\n\s}+\d_{\m\s}\d_{\n\r}-\frac{2}{n}\d_{\m\n}
\d_{\r\s}
\label{identity}
\EN
Similarly one reads the propagators for the ghosts from
the kinetic terms in (\ref{actionghosts}).
 
As mentioned above, adopting this quantization
method one can compute the exact scaling exponents 
of two--dimensional quantum
gravity \cite{polzam}, \cite{DK}, 
with $\e \rightarrow 0$ in the
strong coupling regime \cite{KKN}.
More precisely, if free matter fields with central charge $c$
are included, the one--loop counterterm leads to a 
renormalization of the gravitational coupling constant
\cite{KKN}
\EQ
\frac{1}{G_0}= \m^{-2\e} \left( \frac{1}{G}-\frac{1}{3}~
\frac{c-25}{\e} \right)
\EN
where $\m$ is the renormalization mass. In the 
strong coupling limit , i.e. $\mid G\mid \gg \mid \e \mid $,
one has
\EQ
\frac{1}{G_0}\sim \m^{-2\e}~  \frac{1}{3}~\frac{25-c}{\e}
\label{strong}
\EN

Now the presence of induced dynamical gravity
will counteraffect
matter around.  We concentrate on the case of a bosonic
$\s$--model described by the following action
\EQ
S_M=-\frac{1}{4\p \a} \int d^n x \sqrt{-g}g^{\m\n}\pa_{\m}
\phi^i\pa_{\n}\phi^j G_{ij}(\phi)
\label{actionM}
\EN
We want to study the effects of the gravitational
interaction on the renormalization group flows of the system
and compute the on--shell
$\b$--function.
To this end it is convenient to use the quantum--background 
field method and expand the action in normal coordinates 
\cite{mukhi} 
\bea
&&S_M=-\frac{1}{4\p \a} \int d^nx \sqrt{-g}g^{\m\n}
\left[ \pa_{\m}\phi^i\pa_{\n}\phi^j G_{ij}(\phi) \right. \nonumber\\
&&~~~~~~~~~~~~~~~~~~~~\nonumber \\
&&~~~~~~~~+2D_{\m}\xi^a\pa_{\n}\phi^iE_{ia}(\phi)
+D_{\m}\xi^aD_{\n}\xi^b\d_{ab}-\pa_{\m}\phi^i\pa_{\n}\phi^j
R_{iajb}(\phi)\xi^a\xi^b\nonumber\\
&&~~~~~~~~~~~\nonumber \\
&&~~~~~~~~-\frac{4}{3}\pa_{\m}\phi^i\xi^a\xi^b D_{\n}\xi^c
R_{caib}(\phi)-\frac{1}{3}\pa_{\m}\phi^i\pa_{\n}\phi^j\xi^a
\xi^b\xi^c D_c R_{iajb}(\phi)\nonumber\\
&&~~~~~~~~~~~~~~~~~\nonumber \\
&&~~~~~~~~-\frac{1}{3}D_{\m}\xi^a \xi^b D_{\n}\xi^c\xi^d
R_{abcd}(\phi)-\frac{1}{2}\xi^a\xi^b\xi^cD_{\m} \xi^d
\pa_{\n}\phi^i D_c R_{daib}(\phi) \nonumber\\
\nonumber \\
&&~~~~~~~~\left. -\frac{1}{12}\pa_{\m}\phi^i\pa_{\n}\phi^j
\xi^a\xi^b\xi^c\xi^d(D_aD_b R_{icjd}(\phi)-4 R^k_{~aib}(\phi) 
R_{kcjd}(\phi))
+ O(\xi^5)\right]
\label{coordnorm}
\ena
where $\xi^a$ are coordinates in the tangent frame and 
$D_{\m}\xi^a=\pa_{\m}\xi^a+\pa_{\m}\phi^i\o_i^{~ab}(\phi)~\xi_b$,
being $E_{ia}$ the vielbein and
$\o_{iab}$ the spin--connection of the target space. 
The propagator of the
quantum fields $\xi^a$ is
\EQ
\langle\xi^a \xi^b\rangle =2\p\a ~ \frac{\d^{ab}}{q^2}
\EN
while the gravity--matter interaction vertices can be obtained from
(\ref{coordnorm}) using the quantum--background
expansion
\EQ
\sqrt{-g}g^{\m\n}=\sqrt{-\hat{g}}~(e^{-h})^{\m}_{~\r}~
\hat{g}^{\r\n} e^{\Phi}
\label{GMcoupling}
\EN
  
The material we have briefly summarized gives the relevant
informations which are needed in order to compute the dressed
$\b$--function in the semiclassical limit 
$c\rightarrow -\infty$. In this approximation it suffices to
consider contributions with only one quantum gravity
line, i.e. only one $G_0$ factor. More precisely
in this limit we have (see eq. (\ref{strong}))
\EQ
G_0\sim -\frac{3}{c}~\e
\label{semiclass}
\EN 
Therefore in what follows
we completely disregard the ghost couplings and
the quantum gravity self--interactions. Our general
strategy consists in identifying the Feynman
diagrams which give rise to divergent integrals,
remove infrared divergences, subtract ultraviolet
subdivergences and extract the overall divergence. 
Dimensional
regularization is used to evaluate divergent integrals.
Infrared infinities are completely removed by an appropriate 
choice of the subtraction procedure which is rather easily implemented
at higher loops \cite{CT}. Subtraction of subdivergences corresponding
to lower--loop renormalizations insures automatic cancellation 
of non--local infinite contributions, so that the 
renormalization counterterms correspond to local 
corrections of the $\s$--model metric \cite{friedan}
\EQ
G_{ij}^B=G_{ij}^R + \sum_{k=1}^{\infty} \frac{1}{\e^k}
T^{(k)}_{ij}(G^R)
\EN
The coefficient of the $1/\e$ pole is the relevant one for 
the $\b$--function computation
\EQ
\b_{ij}(G^R)=2\e G^R_{ij}+2(1+\l\frac{\pa~}{\pa\l})
T^{(1)}_{ij}(\l^{-1}G^R)|_{\l=1}
\label{defbeta}
\EN
In the next
section we review the results at the one--loop level
with special emphasys on the infrared subtraction procedure.

\sect{One--loop dressing and infrared divergences}

The perturbative calculation of the $\b$--function
for the $\s$--model in
(\ref{actionM}), in the absence of dynamical gravity,
has been performed up to several loop orders and we refer the
reader to the relevant literature for a review of the basic, 
standard methods and techniques \cite{friedan}, \cite{mukhi},
\cite{review}.

The one--loop contribution is obtained from the ultraviolet
divergence of the
diagram in fig. 1. It produces the following structure
in the effective action
\EQ
\G^{(1)}_{\infty}=\frac{1}{8\p\e}\int d^nx \sqrt{-\hat{g}}~
\hat{g}^{\m\n} \pa _{\m}\phi^i\pa_{\n}\phi^j R_{ij}
\label{beta1}
\EN
with the Ricci tensor defined as $R_{kl}\equiv R^i_{~kil}$. 

Radiative gravitational corrections to leading order in the
semiclassical limit \cite{TKS} are computed from diagrams 
with one $h_{\m\n}$ or one $\Phi$ insertion.  Both gravity
propagators contain a $G_0$ factor; thus from eqs. (\ref{propagators}),
(\ref{semiclass})
we have that the $h_{\m\n}$ line is $O(\e)$, while
the $\Phi$ line is $O(\e^2)$. This has to be kept in mind when
operating subtractions of subleading divergences and
isolating the $1/\e$ divergences from the relevant integrals.

The gravity dressing of the one--loop $\s$--model $\b$--function 
is determined
from the Feynman diagrams in fig. 2, where the dashed lines denote
either the $h_{\m\n}$ or the $\Phi$ propagator. Other divergent graphs
containing at least one tadpole loop are of no interest here: 
indeed, after 
subtraction of subdivergences,they do not produce $1/\e$ poles
and thus do not contribute to the $\b$--function. For this 
same reason
divergent diagrams with at least one tadpole will be discarded
also at higher--loop level.
 
The evaluation of the
leading divergence of the various contributions from fig. 2
is rather simple. Even the removal of infrared divergences
does not present technical difficulties 
to this loop order: it can be implemented
easily using different but equivalent methods. However, at 
higher--loop level, we have
found advantageous to adopt an infrared regularization \cite{CT}
which essentially parallels the BPHZ subtraction method for 
ultraviolet divergences. The idea is to use a procedure which
completely removes infrared infinities from any given
integral. Consequently we are free to modify the
various diagrams (even changing their infrared behaviour),
as long as we do not alter their ultraviolet
divergent nature.
This allows for example to evaluate all the graphs at zero
external momenta and/or route an external momentum through
the graph in some convenient way \cite{Vlad}, \cite{CKT}. 
In so doing the infrared 
divergences will be rearranged, but we need not worry since 
in the end they will all be removed.

We illustrate this method on a simple 
example and
compute in detail the contribution from the diagram in fig. 2a,
with the dashed line denoting the $h_{\mu \nu}$ propagator. 
One obtains a contribution proportional to 
$ h^{\mu \r \nu \sigma} ~I_{\mu \nu \r \sigma} $  
with $I_{\mu \nu \r \sigma}$ given by the following integral
\EQ
I_{\mu \nu \r \sigma} = 
G_0 \int d^nq~ d^nr~ \frac{q_{\m}q_{\n}(q-r)_{\r} (q-r)_{\s}}
{(q^2)^2 ~(q-r)^2 ~r^2}
\label{int1}
\EN
In order to keep the notation manageable in the main text,
we use the
convention of dropping factors of $(2 \p)^{-n}$ for each
loop integral, with the understanding that at the end we
will have to reinsert a factor $(4\p)^{-\frac{n}{2}}$
for each loop. The reader can find in the Appendices the relevant
formulas with all the factors spelled out.
Using the first IR subtraction formula listed 
in Appendix A, eq. (\ref{a1}), one easily performs the $r$--integral 
\bea
&&\int d^nr \frac{(q-r)_{\r}(q-r)_{\s}}{(q-r)^2 r^2}
\rightarrow \int d^nr \frac{(q-r)_{\r}(q-r)_{\s}}{(q-r)^2 r^2}
+\frac{1}{\e}~\frac{q_{\r}q_{\s}}{q^2}= \nonumber\\
&&~~~~~=(1+2\e)\left\{ \frac{1}{2\e}~\frac{\d_{\r\s}}{(q^2)^{\e}}
-\frac{1-\e}{\e}~\frac{q_{\r}q_{\s}}{(q^2)^{1+\e}}
\right\} + \frac{1}{\e}~\frac{q_{\r}q_{\s}}{q^2}+O(\e)
\ena
Then the integration over the $q$--variable leads to
\EQ
I_{\mu \nu \r \sigma} \sim G_0 \left[
\frac{1}{16\e^2}(1+\frac{\e}{2})3\d_{(\m\n}\d_{\r\s)}
+\frac{1}{8\e^2}(1+3\e)\d_{\m\n} \d_{\r\s} \right]
\label{integral}
\EN
where, in order to handle infrared subtractions, use of (\ref{a2}) and 
(\ref{a3}) has been made.
Since $G_0 = O(\e)$ only the second order poles in eq. (\ref{integral})
are relevant. 
Now we must subtract subleading ultraviolet divergences. 
The only divergent sub--diagram is the one which does not contain
the gravity propagator (the loop containing the 
$h_{\mu \nu}$
line gives $O(G_0/\e)\sim O(1)$  finite contributions).
Therefore we concentrate on the $q$--loop. Setting
$r=0$ we find
\EQ
\left[ \int d^nq \frac{q_{\m} q_{\n} q_{\r} q_{\s}}{(q^2)^3} \right]_{div}
=\frac{1}{8\e} 3\d_{(\m\n}\d_{\r\s)}
\EN
Once again we have used eq. (\ref{a3}). 
Thus to eq. (\ref{integral}) one must add
the following divergent contribution 
\EQ
-\frac{1}{8\e} 3\d_{(\m\n}\d_{\r\s)}G_0\int d^nr\frac{1}{r^2}=
-\frac{1}{8\e^2} 3\d_{(\m\n}\d_{\r\s)}G_0
\EN
Finally, using the identities in Appendix A, namely 
eqs. (\ref{a10}, \ref{a11}), one finds
\EQ
h^{\mu \r \nu \sigma} ~I_{\mu \nu \r \sigma} \rightarrow 
G_0\left[-\frac{1}{16\e^2}8+\frac{1}{8\e^2}4\right] = 0
\EN
so that the diagram in fig. 2 with an $h_{\m\n}$ line does not
contribute to the $\b$--function.

One easily reaches the same conclusion when the gravity 
insertion corresponds to a $\Phi$ line. In this case the propagator 
is $O(\e^2)$
so that it
cancels the $1/\e^2$ poles from the loop integrations and no divergent 
contributions survive.

The integrals corresponding to the diagrams in fig. 2b, 2c
can be evaluated in similar manner. It is straightforward
to show that both contributions separately vanish so that, as
anticipated in the introduction, one finds \cite{TKS}
\EQ
\b^{(1)}_g =0
\EN
At one--loop order in the $\s$--model fields
the gravitational dressing is simply given by the multiplicative factor
from the cosmological constant renormalization. Now we turn to the
next order in perturbation theory.

\sect{Gravitational corrections to two--loop matter}

We consider the gravitational dressing of the two--loop matter 
$\b$--functions. 
The standard two--loop $\b$--function without gravity is obtained 
from the divergences
of the diagram in fig. 3. The corresponding contribution to the 
effective action is given by
\EQ
\G^{(2)}_{\infty}= \frac{\a}{32\p\e} \int d^nx 
\sqrt{-\hat{g}}\hat{g}^{\m\n}\pa_{\m}\phi^i \pa_{\n}\phi^j
R_{iklm}R_j^{~klm}
\EN
In order to evaluate corrections due to the presence of
dynamical gravity in the semiclassical
limit we need consider all possible two--loop matter graphs with 
the insertion of
one $h_{\mu \nu}$ or one $\Phi$ propagator. The corresponding, 
various contributions can be grouped on the basis of the 
background structures they end up being proportional to.
In order to write the answer in terms of independent tensors
we make use of the following relation
\EQ
D^a D^b R_{iajb} = -R_i^{~a} R_{ja} + R_{iajb} R^{ab} + D^a D_a R_{ij} 
- \frac12 D_i D_j R
\label{cyclic}
\EN
Moreover we drop the term $D_iD_jR$ since it gives rise to 
a contribution in the effective action which vanishes
on--shell. In so doing we can assemble the total sum
of terms relevant for the $\b$--function calculation
in the form
\EQ
\frac{\a}{12\pi} \frac{G_0}{\e^2} 
\int d^nx \sqrt{-\hat{g}}\hat{g}^{\m\n}
\pa_{\m}\phi^i \pa_{\n}\phi^j\left( a_1 R_{iklm}R_j^{~klm}
+a_2 R_{ikjl}R^{kl} +a_3 R_{ik}R_{j}^{~k} +a_4 D^k D_k R_{ij}\right)
\label{totalsum}
\EN
with the numerical coefficients $a_i$ to be determined by
the explicit evaluation of the relevant Feynman diagrams.
From the action in eq. (\ref{coordnorm}) it is easy to see how 
the structures in eq. (\ref{totalsum}) are produced from 
graphs which combine in various ways
the quantum--background vertices. 
In figs. 4, 5, 6  we have drawn all the interesting topologies:
they give
 non--vanishing contributions respectively to the $a_1$, $a_2$, $a_3$
coefficients, while the structure proportional to $a_4$ is produced
by a diagram like the one in fig. 6a.
As mentioned earlier graphs containing at least
one tadpole have not been included since they do not contribute 
to the $1/\e$ pole.

As an example of three--loop calculation we give details
of the evaluation 
of the diagram in fig. 4a with a $h_{\mu \nu}$ gravity
correction. To start with there are four different integrals
associated with this diagram, according to the four distinct 
ways to contract the fields. 
By integration by parts they can be reduced to the two structures 
schematically shown in
fig. 7, where the arrows denote derivatives acting on the 
corresponding propagators. 
We concentrate on the calculation of the
loop integrals for the diagram in fig. 7a. Inserting the
appropriate combinatoric factor, 
the contribution from this diagram is
\EQ
\frac{9\a}{\pi} ~h^{\r \sigma \t \pi} ~J_{\mu \nu \r \sigma \t \pi}
~\hat{g}^{\mu \g} \hat{g}^{\nu \d} ~\partial_{\g} \phi^i \partial_{\d} \phi^j
~R_{iklm} R_j^{~klm} 
\label{diagr4a}
\EN
where
\EQ
J_{\mu \nu \r \sigma \t \pi} \equiv G_0 \int d^nk~ d^nq~ d^nr~ 
\frac{k_{\m}(k-q)_{\n}q_{\r}(q-r)_{\s}
q_{\t}(q-r)_{\p}}{k^2(k-q)^2(q^2)^2(q-r)^2 r^2}
\label{int2}
\EN
We can easily perform the $k$--integration in eq. (\ref{int2})
using the results (\ref{b3}, \ref{b4}) listed in Appendix B.
We obtain
\bea
J_{\mu \nu \r \sigma \t \pi}&=&
G_0 (1+2\e) \left\{ \frac{1}{2\e} \d_{\m\n} \int d^nq d^nr
\frac{q_{\r}(q-r)_{\s}q_{\t}(q-r)_{\p}}{(q^2)^{2+\e}(q-r)^2
r^2} \right. \nonumber\\
&&~~~~~~~~~~~\left. -\int d^nq d^nr \frac{q_{\m} q_{\n} 
q_{\r}(q-r)_{\s}q_{\t}(q-r)_{\p}}{(q^2)^{3+\e} (q-r)^2 r^2}
\right\}
\ena
As a second step we evaluate the $r$--integral. We make
use of eqs. (\ref{b2}, \ref{b3}, \ref{b4}) for the
momentum integrals and of 
the relation in (\ref{a1}) for the removal of the infrared
divergence 
\bea
J_{\mu \nu \r \sigma \t \pi} &=&
G_0 (1+4\e)\left\{ \frac{1}{4\e^2}\d_{\m\n} \d_{\s\p}\int
d^nq\frac{q_{\r}q_{\t}}{(q^2)^{2+2\e}}\right.\nonumber\\
&~&~-\frac{1-\e}{2\e^2} \d_{\m\n}\int d^nq
\frac{q_{\r}q_{\s}q_{\t}q_{\p}}{(q^2)^{3+2\e}}-
\frac{1}{2\e}\d_{\s\p}\int d^nq
\frac{q_{\m}q_{\n}q_{\r}q_{\t}}{(q^2)^{3+2\e}}\nonumber\\
&~&~\left. +\frac{1-\e}{\e} \int d^nq \frac{q_{\m}q_{\n}
q_{\r}q_{\s}q_{\t}q_{\p}}{(q^2)^{4+2\e}}\right\} \nonumber\\
&~&~+(1+2\e) \left\{ \frac{1}{2\e^2} \d_{\m\n}\int d^nq
\frac{q_{\r}q_{\s}q_{\t}q_{\p}}{(q^2)^{3+\e}}-
\frac{1}{\e} \int d^nq \frac{q_{\m}q_{\n}
q_{\r}q_{\s}q_{\t}q_{\p}}{(q^2)^{4+\e}}\right\}
\ena
Finally the $q$--integration is readily done with the help of
the infrared prescriptions in eqs. (\ref{a2}, \ref{a3}, \ref{a4}). 
The result is
\bea
&& J_{\mu \nu \r \sigma \t \pi}= G_0 \left[
\frac{1}{24\e^3}(1+5\e)\d_{\m\n}\d_{\s\p}\d_{\r\t}
+\frac{1}{96\e^3} (1+\frac{3}{2}\e)\d_{\m\n}3\d_{(\r\s}\d_{\t\p)}
\right. \nonumber\\
&~&~~~~~~~~~~\left. -\frac{1}{48\e^2}\d_{\s\p}3\d_{(\m\n} \d_{\r\t)}-
\frac{1}{288\e^2}15\d_{(\m\n}\d_{\r\s}\d_{\t\p)} \right]
\label{int3}
\ena
Now we consider UV subdivergences that, if present, we need
subtract from the
expression in (\ref{int3}).  At one loop the only subdivergence 
comes from one of the sub--diagrams 
which do not contain the gravity line, i.e. the one
given by the $k$--integral. Power counting would single out another
potential subdivergence associated to the $r$--integral:
in fact this loop, 
containing the $h_{\m\n}$ gravity propagator, 
is effectively $O(G_0/\e)$ and thus it gives rise to a finite contribution.
The divergence from the $k$--integral is $\delta_{\mu \nu}(2\e)^{-1}$ 
and the corresponding one--loop subtraction 
is given by  
\bea
&&-\frac{\d_{\m\n}}{2\e} G_0 \int d^nq d^nr \frac{q_{\r}(q-r)_{\s}
q_{\t}(q-r)_{\p}}{(q^2)^2(q-r)^2 r^2}= \nonumber\\
&&~~~~~~~~~=-\frac{1}{16\e^3}(1+3\e)\d_{\m\n}\d_{\s\p}
\d_{\r\t}-\frac{1}{32\e^3}(1+
\frac{1}{2}\e)\d_{\m\n}3\d_{(\r\s}\d_{\t\p)}
\label{int4}
\ena
At two loops the UV divergence which must be subtracted is the one
associated to the $k$ and $q$ integrals. Setting $r=0$
we write
\bea
&&\left[ \int d^nk d^nq \right]_{div} =  
\int d^nk d^nq \frac{k_{\m}(k-q)_{\n}q_{\r}q_{\s}
q_{\t}q_{\p}}{k^2(k-q)^2(q^2)^3}= \nonumber\\
&&=(1+2\e) \left\{ \frac{1}{2\e} \d_{\m\n} \int
d^nq \frac{q_{\r}q_{\s}
q_{\t}q_{\p}}{(q^2)^{3+\e}} -\int d^nq \frac{q_{\m} q_{\n}q_{\r}q_{\s}
q_{\t}q_{\p}}{(q^2)^{4+\e}} \right\}
\ena
Again the $q$--integral is trivially performed using the
relations (\ref{a3}, \ref{a4}) in Appendix A.
After subtraction of its own one--loop subdivergence 
from the $k$--loop, it finally gives
\EQ
\left[\int d^nk d^nq\right]_{sub}= 
-\frac{1}{32\e^2}(1-\frac{1}{2}\e) \d_{\m\n}
3\d_{(\r\s} \d_{\t\p)} -\frac{1}{96\e} 15\d_{(\m\n}\d_{\r\s}
\d_{\t\p)}
\EN
In sum, the two--loop subtraction to be performed in 
eq. (\ref{int3}) is given by
\bea
&& \left[ \frac{1}{32\e^2}(1-\frac{1}{2}\e) \d_{\m\n}
3\d_{(\r\s} \d_{\t\p)} +\frac{1}{96\e} 15\d_{(\m\n}\d_{\r\s}
\d_{\t\p)} \right] G_0 \int d^nr \frac{1}{r^2} =
\nonumber \\
&=& G_0 \left[ \frac{1}{32\e^3}(1-\frac{1}{2}\e) \d_{\m\n}
3\d_{(\r\s} \d_{\t\p)} +\frac{1}{96\e^2} 15\d_{(\m\n}\d_{\r\s}
\d_{\t\p)} \right]
\label{int5}
\ena
(We note that the other two--loop integral in the $k$ and $r$
variables is finite after subtraction of its one--loop $k$
subdivergence.)
 
Adding the results in eqs. (\ref{int3}), (\ref{int4}), (\ref{int5}) 
and contracting the various structures with $h^{\r\s\t\pi}$
as indicated in eq. (\ref{diagr4a}), one can check that the contribution
from the graph in fig. 7a with a gravity $h_{\m\n}$ correction
is zero.
The diagram in fig. 7b can be evaluated following the same procedure. 
Its contribution to the $a_1$ coefficient in eq. (\ref{totalsum}) is
$1/4$. 

In an analogous way one can compute the $1/\e$ divergences
from the two diagrams
in fig. 7 when a $\Phi$ propagator is inserted. Again the first graph 
gives a vanishing contribution, while the second one exactly cancels 
the corresponding $h_{\mu \nu}$ correction. 

We are now ready to list the results for all the diagrams contributing
to the expression in eq. (\ref{totalsum}). 
Each graph receives several contributions stemming from the 
different ways of arranging derivatives at the 
vertices. Integration by parts is used repeatedly in order to obtain
an independent set of configurations. 
For each graph we list the total contribution to the coefficients $a_i$,
keeping distinct the correction from the $h_{\m\n}$ and the
$\Phi$ propagator. 

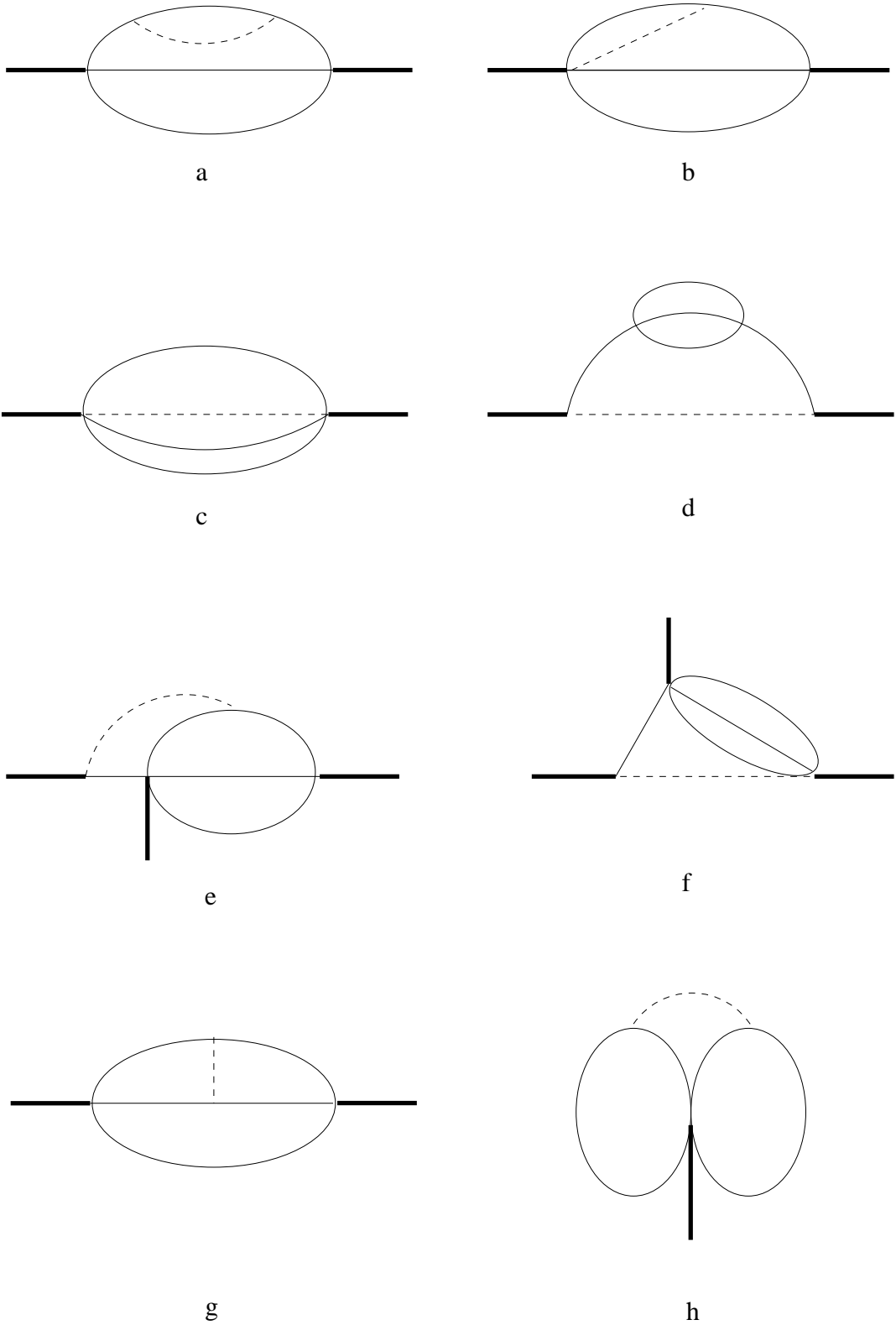
\subsection{Contributions proportional to the Riemann--Riemann structure}

We present here the results for the diagrams in fig. 4 which all 
end up producing a background dependence of the form Riemann--Riemann,
thus contributing to the $a_1$ coefficient in (\ref{totalsum}). 
We have obtained
\bea
~~ &~&~~~~~~~~~~~~~~~~~h_{\m\n}~~~~~~~~~~~~~~~\Phi  \nonumber\\
~~~~&~&~~~~\nonumber\\
4a &:&~~~~~~~~~~~~~~~~~ \frac{1}{4} ~~~~~~~~~~~~~~-\frac{1}{4} \nonumber \\
4b &:&~~~~~~~~~~~~~~~- \frac{5}{2} ~~~~~~~~~~~~~~~~~\frac12    \nonumber \\
4c &:&~~~~~~~~~~~~~~~~ \frac{1}{6} ~~~~~~~~~~~~~~-\frac{1}{6} \nonumber \\
4d &:&~~~~~~~~~~~~~- \frac{13}{12}~~~~~~~~~~~~~~~~\frac{1}{12} \nonumber \\
4e &:&~~~~~~~~~~~~~~~~0           ~~~~~~~~~~~~~~~~~~ 0           \nonumber \\
4f &:&~~~~~~~~~~~~~~~ \frac{13}{6} ~~~~~~~~~~~~~~-\frac{1}{6}  \nonumber \\
4g &:&~~~~~~~~~~~~~~~~ 1~~~~~~~~~~~~~~~~~~\frac{1}{2}  \nonumber \\
4h &:&~~~~~~~~~~~~~~~~0           ~~~~~~~~~~~~~~-\frac{1}{2}  
\ena
The total sum is identically zero
so that the coefficient $a_1$ in eq. (\ref{totalsum}) vanishes.
Notice in particular that a vanishing result is obtained for the 
$h_{\mu \nu}$ and the $\Phi$ contributions separately.

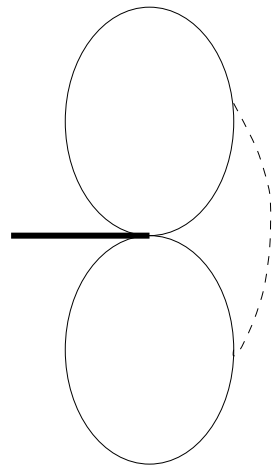
\subsection{Contributions proportional to the Riemann--Ricci structure}

Here we give the results for the diagrams in fig. 5. In this case the
graphs $c$ and $d$ with a $h_{\mu \nu}$ insertion do not contribute, 
since the tensorial structure of the loop--integrals  
vanishes when contracted with $h^{\mu \nu \r \sigma}$ (see Appendix A
eqs. (\ref{a8}--\ref{a14})).
One obtains  
\bea
~~ &~&~~~~~~~~~~~~~~~~h_{\m\n}~~~~~~~~~~~~~~~\Phi  \nonumber\\
~~~~&~&~~~~\nonumber\\
5a &:& ~~~~~~~~~~~~~~-\frac12      ~~~~~~~~~~~~~~~~\frac13 \nonumber \\
5b &:& ~~~~~~~~~~~~~~~~ \frac12          ~ ~~~~~~~~~~~~~~~~ 1      \nonumber \\
5c &:& ~~~~~~~~~~~~~~~~ 0           ~~~~~~~~~~~~~~-1       \nonumber \\
5d &:& ~~~~~~~~~~~~~~~~ 0~~~~~~~~~~~~~~~~~~\frac{1}{6}
\ena
Again the sum of the contributions for the $h_{\m\n}$ field is zero,
but a nonzero $a_2$ coefficient for the $R_{ikjl}R^{kl}$ tensor 
is produced by the $\Phi$ coupling.

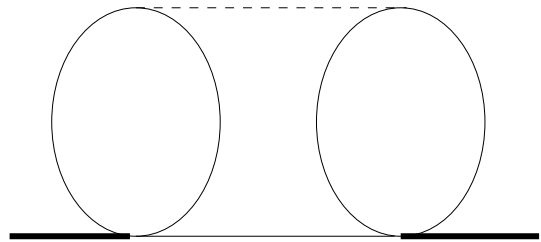
\subsection{Contributions proportional to the Ricci--Ricci structure}

It is easy to verify that the two diagrams in fig. 6 
with a $h_{\mu \nu}$ insertion do not contribute since, as before,
the loop--integrals give a zero result when
multiplied by $h^{\mu \nu \r \sigma}$. We have
\bea
~~ &~&~~~~~~~~~~~~~~~~h_{\m\n}~~~~~~~~~~~~~~~\Phi  \nonumber\\
~~~~&~&~~~~\nonumber\\
6a &:& ~~~~~~~~~~~~~~~~0      ~~~~~~~~~~~~~~~-\frac12 \nonumber \\
6b &:& ~~~~~~~~~~~~~~~~ 0          ~~~~~~~~~~~~~~~~~~ \frac12 
\ena
and the coefficient $a_3$ in eq. (\ref{totalsum}) vanishes.

\subsection{Contributions proportional to the $D^kD_k R_{ij}$ structure}

At this point it is an easy task to evaluate the contribution
to the coefficient $a_4$ in eq. (\ref{totalsum}). Indeed
all the relevant work has already been done and one simply has
to reconsider a diagram with the topology shown in fig. 6a.
Last but not least, it gives
\bea
~~ &~&~~~~~~~~~~~~~~~h_{\m\n}~~~~~~~~~~~~~\Phi  \nonumber\\
~~~~&~&~~~~\nonumber\\
6a &:& ~~~~~~~~~~~~~~~0      ~~~~~~~~~~~~~~~~\frac14
\ena

Now an important comment is in order: all 
the contributions produced at intermediate stages by the
coupling of matter to the gravity $h_{\m\n}$ field in the 
end cancel out completely. This result, expected from a
formulation of gravity in two--dimensional conformal gauge,
provides a nontrivial check of the general approach and of
the actual calculation.  

In conclusion the coefficients $a_2$ and $a_4$ in (\ref{totalsum})
are nonvanishing and a
contribution to the $\b$--function is produced
from gravity radiative corrections.
We discuss the relevance of this result in the last section.  

\sect{Discussion and conclusions}

The final answer for the $\b$--function up to two loops
follows from a straightforward application of the
definition in eq. (\ref{defbeta})
\EQ
\b^{(1)}_{ij} +\b^{(2)}_{ij} = R_{ij}+\frac{\a}{2}[R_{iklm}R_{j}^{~klm}-
\frac{2}{c}(D^kD_k R_{ij}+ 2R_{ikjl} R^{kl})]
\label{betatotal}
\EN
The remaining gravitational dressing is then obtained multiplying
the above expression by the prefactor in the r.h.s. of 
eq. (\ref{beta}). As already mentioned this factor tells 
how the renormalization
mass $\m$ changes with respect to the physical scale, which 
is determined in turn by the renormalization of the cosmological term.
It is clear that while at one loop all the informations
about the influence of the gravitational fields on the
$\s$--model matter system are contained just in this
multiplicative factor, this fails to be true at the two--loop
level because of the presence of the last contribution in
(\ref{betatotal}). Thus we have found for the $\s$--model
the same pattern as the one suggested
for a system with just one coupling constant \cite{dorn}:
beyond leading order in perturbation theory the  renormalization 
group trajectories, and consequently the fixed points 
are modified significantly by the interaction with the dynamical
gravity.

We have performed our calculation using a particular regularization
procedure and we have to ask ourselves how the scheme dependence might
affect the result we have presented. The idea is to
take into account conventional subraction ambiguities which would
arise from finite
subtractions proportional to the one--loop counterterms.
Since the quantum--background splitting we have used is non--linear,
some care is required 
\cite{stelle}: one needs compute the complete set of
one--loop counterterms with quantum matter and gravity fields.
One can check explicitly that all the counterterm insertions
containing a gravity field ( both the ones with $h_{\m\n}$ and 
the ones with $\Phi$ ) do not contribute to the $\b$--function, so that
the net effect at two loops is  the addition to (\ref{betatotal})
of a term proportional to the following expression
\EQ
R_{ik}R_j^{~k}-R_{ikjl}R^{kl}-\frac{1}{2}D^kD_kR_{ij}
\label{scheme}
\EN
This is nothing but the usual dependence on the regularization scheme
of the flat two-loop $\b$--function, with no extra corrections 
due to gravity.  
One might conclude asserting that the modification
of the renormalization group flows is not an artifact
of the calculation. It reflects instead the new physics
induced by the curved quantum two--dimensional spacetime.

In Ref. \cite{GZ} the exact dressing of one--loop 
$\s$--model $\b$--functions has been obtained for
both $N=1$ and $N=2$ supersymmetric theories coupled to
induced supergravity. More precisely it has been shown that
for the $N=1$ case
 \EQ
\b^{(1)}_G=\frac{\k+\frac{3}{2}}{\k+1}~\b^{(1)}_0
\EN
with
\EQ
\k+\frac{3}{2}=\frac{1}{8}\left[c-5-\sqrt{(1-c)(9-c)}\right]
\EN
while for the $N=2$ theory no supergravity dressing is produced
for the one--loop $\b$--function.
The corresponding analysis at the two--loop level in the matter
fields is presently under consideration.

\vskip 10pt
{\bf Acknowledgements}: D. Zanon thanks M. Grisaru for useful 
conversations. This work has been partially supported by
grants no. SC1--CT92--0789 and no. CEE--CHRX--CT92--0035. 

\vskip 20pt
\appendix
\sect{Infrared counterterms and useful formulae}

As emphasized earlier,
in our work we had to deal with the typical infrared
divergences of massless fields in two dimensions. We have used
an infrared regularization procedure which amounts to regulate
directly every infrared divergent factor \cite{CT}. For example for each
$1/p^2$ term one introduces a counterterm
$a\d^{(2)}(p)$ in such a fashion that
\EQ
\int \frac{d^n p}{(2\pi)^n} f(p)\left[ \frac{1}{p^2}-a\d^{(2)}(p)\right]
\label{test}
\EN
is finite for any test function $f(p)$ which vanishes at
infinity. Choosing in particular $f(p)=1/(p^2+m^2)$ one 
immediately determines $a=-\pi/\e$. Similarly one obtains 
\bea
\frac{1}{(p^2)^{1+(\l-1)\e}} &\rightarrow & \frac{1}{(p^2)^{1+(\l-1)\e}}+
\frac{\pi}{\l\e}\d^{(2)}(p) \label{a1}\\
\frac{p_{\m}p_{\n}} {(p^2)^{2+(\l-1)\e}} &\rightarrow & \frac{p_{\m}p_{\n}}
{(p^2)^{2+(\l-1)\e}}+\pi \frac{\d_{\m\n}} {2\l\e(1-\e)}\d^{(2)}(p) \label{a2}\\
\frac{p_{\m}p_{\n}p_{\r}p_{\s}} {(p^2)^{3+(\l-1)\e}} &\rightarrow &
\frac{p_{\m}p_{\n}p_{\r}p_{\s}} {(p^2)^{3+(\l-1)\e}} + 
\pi \frac{3\d_{(\m\n}\d_{\r\s)}} {4\l\e(1-\e)(2-\e)}\d^{(2)}(p) \label{a3}\\
\frac{p_{\m}p_{\n}p_{\r}p_{\s}p_{\t}p_{\p}} {(p^2)^{4+(\l-1)\e}}
&\rightarrow & \frac{p_{\m}p_{\n}p_{\r}p_{\s}p_{\t}p_{\p}}
{(p^2)^{4+(\l-1)\e}} +
\pi \frac{15\d_{(\m\n}\d_{\r\s}\d_{\t\p)}} {8\l\e(1-\e)(2-\e)(3-\e)}
\d^{(2)}(p)
\label{a4}
\ena
The ones listed above are all the infrared counterterms that we 
have used in the evaluation of our diagrams.

In the course of the calculation we made repeated use of some identities
that we list here for the convenience of the reader 
\bea
\d^{\r\s}3\d_{(\m\n}\d_{\r\s)} &=& 4(1-\frac{1}{2}\e)\d_{\m\n}
\label{a5}\\
\d^{\t\p}15\d_{(\m\n}\d_{\r\s}\d_{\t\p)} &=& 6(1-\frac{1}{3}\e)
3\d_{(\m\n}\d_{\r\s)} \label{a6}\\
3\d^{(\m\n}\d^{\r\s)}3\d_{(\m\n}\d_{\r\s)} &=& 24(1-\frac{3}{2}\e)
\label{a7}
\ena
In addition, with the definition in eq. (\ref{identity})
one also obtains
\bea
\d_{\m\n}h^{\m\n\r\s} &=& 0 \label{a8}\\
\d_{\m\r}h^{\m\n\r\s} &=& 2(1-\frac{3}{2}\e)\d^{\n\s} \label{a9}\\
\d_{\m\r}\d_{\n\s}h^{\m\n\r\s} &=& 4(1-\frac{5}{2}\e) \label{a10}\\
3\d_{(\m\n}\d_{\r\s)}h^{\m\n\r\s} &=& 8(1-\frac{5}{2}\e) \label{a11}\\
3\d_{(\m\n}\d_{\r\s)}h^{\m'\m\r\s} &=& 4(1-\frac{3}{2}\e)\d^{\m'}_{\n}
\label{a12}\\
3\d_{(\m'\n'}\d_{\m\r)}\d_{\n\s}h^{\m\n\r\s} &=& 8(1-2\e)\d_{\m'\n'}
\label{a13}\\
15\d_{(\m'\n'}\d_{\m\n}\d_{\r\s)}h^{\m\n\r\s} &=& 24(1-\frac{11}{6}\e)
\d_{\m'\n'}
\label{a14}
\ena

\sect{The basic integrals}

We collect in this appendix a list of 
momentum integrals we have encountered in our calculation. We 
have used dimensional regularization in the
so called G--scheme \cite{CKT} which is a form
of modified minimal subtraction.
By introducing a factor
$\G(1-\e)(4\p)^{-\e}$ for each loop integral, it  leads to the automatic 
cancellation of irrelevant factors of $ln4\p$, $\g_E$ and $\z(2)$. 
We have obtained the following results:
\bea
&& \G(1-\e)(4\p)^{-\e}\int \frac{d^nk}{(2\p)^n} 
\frac{1}{(k^2)^{(\l-1)\e}(k-r)^2}=\nonumber\\
&& =\frac{1}{4\p}
\frac{\G(1-\e)\G(\l\e)\G(-\e)\G(1-\l\e)} {\G((\l-1)\e)\G(1-(\l+1)\e)}
\frac{1}{(r^2)^{\l\e}}
\label{b1}\\
&&~\nonumber\\
&& \G(1-\e)(4\p)^{-\e}\int \frac{d^nk}{(2\p)^n} 
\frac{1}{(k^2)^{1+(\l-1)\e}(k-r)^2}=\nonumber\\
&& =\frac{1}{4\p}
\frac{\G(1-\e)\G(1+\l\e)\G(-\e)\G(-\l\e)} {\G(1+(\l-1)\e)\G(-(\l+1)\e)}
\frac{1}{(r^2)^{1+\l\e}}
\label{b2}\\
&&~\nonumber\\
&& \G(1-\e)(4\p)^{-\e}\int \frac{d^nk}{(2\p)^n}
 \frac{k_{\m}} {(k^2)^{1+(\l-1)\e}(k-r)^2}=\nonumber\\
&& =\frac{1}{4\p}
\frac{\G(1-\e)\G(1+\l\e)\G(-\e)\G(1-\l\e)} {\G(1+(\l-1)\e)\G(1-(\l+1)\e)}
\frac{r_{\m}} {(r^2)^{1+\l\e}}
\label{b3}\\
&&~\nonumber\\
&& \G(1-\e)(4\p)^{-\e}\int \frac{d^nk}{(2\p)^n}
 \frac{k_{\m}k_{\n}} {(k^2)^{1+(\l-1)\e}(k-r)^2}=\nonumber\\
&& =\frac{1}{4\p}
\frac{\G(1-\e)}{\G(1+(\l-1)\e)\G(2-(\l+1)\e)}\frac{1}{(r^2)^{\l\e}}
\left[\G(1+\l\e)\G(-\e)\G(2-\l\e)\frac{r_{\m}r_{\n}} {r^2}
\right.\nonumber\\
&~&~~~ \left.+\shalf\G(\l\e)\G(1-\e)\G(1-\l\e)\d_{\m\n}\right]
\label{b4}\\
&&~\nonumber\\
&&\G(1-\e)(4\p)^{-\e} \int \frac{d^nk}{(2\p)^n}
 \frac{k_{\m}k_{\n}} {(k^2)^{2+(\l-1)\e}(k-r)^2}=\nonumber\\
&& =\frac{1}{4\p}
\frac{\G(1-\e)}{\G(2+(\l-1)\e)\G(1-(\l+1)\e)}\frac{1}{(r^2)^{1+\l\e}}
\left[\G(2+\l\e)\G(-\e)\G(1-\l\e)\frac{r_{\m}r_{\n}} {r^2}
\right.\nonumber\\
&~&~~~ \left.+\shalf\G(1+\l\e)\G(1-\e)\G(-\l\e)\d_{\m\n}\right]
\label{b5}\\
&&~\nonumber\\
&& \G(1-\e)(4\p)^{-\e}\int \frac{d^nk}{(2\p)^n}
 \frac{k_{\m}k_{\n}k_{\r}} {(k^2)^{2+(\l-1)\e}(k-r)^2}=
\nonumber\\
&& =\frac{1}{4\p}
\frac{\G(1-\e)}{\G(2+(\l-1)\e)\G(2-(\l+1)\e)}\frac{1}{(r^2)^{1+\l\e}}
\left[\G(2+\l\e)\G(-\e)\G(2-\l\e)\frac{r_{\m}r_{\n}r_{\r}} {r^2}
\right.\nonumber\\
&~&~~~ \left.+\shalf\G(1+\l\e)\G(1-\e)\G(1-\l\e)3\d_{(\m\n}r_{\r)}
\right]
\label{b6}\\
&&~\nonumber\\
&& \G(1-\e)(4\p)^{-\e}\int \frac{d^nk}{(2\p)^n}
 \frac{k_{\m}k_{\n}k_{\r}k_{\s}} {(k^2)^{2+(\l-1)\e}(k-r)^2}=
\nonumber\\
&& =\frac{1}{4\p}
\frac{\G(1-\e)}{\G(2+(\l-1)\e)\G(3-(\l+1)\e)}\frac{1}{(r^2)^{\l\e}}
\left[\G(2+\l\e)\G(-\e)\G(3-\l\e)\frac{r_{\m}r_{\n}r_{\r}r_{\s}} 
{(r^2)^2}\right.\nonumber\\
&& \left.+\shalf\G(1+\l\e)\G(1-\e)\G(2-\l\e)
\frac{6\d_{(\m\n}r_{\r}r_{\s)}} {r^2}+\frac{1}{4}\G(\l\e)
\G(2-\e)\G(1-\l\e)3\d_{(\m\n}\d_{\r\s)}\right]
\label{b7}\\
&&~\nonumber\\
&& \G(1-\e)(4\p)^{-\e}\int \frac{d^nk}{(2\p)^n}
 \frac{k_{\m}k_{\n}k_{\r}k_{\s}} {(k^2)^{3+(\l-1)\e}(k-r)^2}=
\nonumber\\
&& =\frac{1}{4\p}
\frac{\G(1-\e)}{\G(3+(\l-1)\e)\G(2-(\l+1)\e)}\frac{1}{(r^2)^{1+\l\e}}
\left[\G(3+\l\e)\G(-\e)\G(2-\l\e)\frac{r_{\m}r_{\n}r_{\r}r_{\s}} 
{(r^2)^2}\right.\nonumber\\
&& \left.+\shalf\G(2+\l\e)\G(1-\e)\G(1-\l\e)
\frac{6\d_{(\m\n}r_{\r}r_{\s)}} {r^2}+
\frac{1}{4}\G(1+\l\e)\G(2-\e)\G(-\l\e)
3\d_{(\m\n}\d_{\r\s)}\right]
\label{b8}\\
&&~\nonumber\\
&& \G(1-\e)(4\p)^{-\e}\int \frac{d^nk}{(2\p)^n}
 \frac{k_{\m}k_{\n}k_{\r}k_{\s}k_{\t}} 
{(k^2)^{3+(\l-1)\e}(k-r)^2}=\nonumber\\
&& =\frac{1}{4\p}
\frac{\G(1-\e)}{\G(3+(\l-1)\e)\G(3-(\l+1)\e)}\frac{1}{(r^2)^{1+\l\e}}
\left[\G(3+\l\e)\G(-\e)\G(3-\l\e)\frac{r_{\m}r_{\n}r_{\r}r_{\s}
r_{\t}} {(r^2)^2}\right.\nonumber\\
&& +\shalf\G(2+\l\e)\G(1-\e)\G(2-\l\e)
\frac{10\d_{(\m\n}r_{\r}r_{\s}r_{\t)}} {r^2} \nonumber \\
&& \left. +\frac{1}{4}\G(1+\l\e)\G(2-\e)\G(1-\l\e)
15\d_{(\m\n}\d_{\r\s}r_{\t)}\right]
\label{b9}\\
&&~\nonumber\\
&&\G(1-\e)(4\p)^{-\e} \int \frac{d^nk}{(2\p)^n}
 \frac{k_{\m}k_{\n}k_{\r}k_{\s}k_{\t}k_{\p}} 
{(k^2)^{3+(\l-1)\e}(k-r)^2}=\nonumber\\
&& =\frac{1}{4\p}
\frac{\G(1-\e)}{\G(3+(\l-1)\e)\G(4-(\l+1)\e)}\frac{1}{(r^2)^{1+\l\e}}
\left[\G(3+\l\e)\G(-\e)\G(4-\l\e)\frac{r_{\m}r_{\n}r_{\r}r_{\s}
r_{\t}r_{\p}} {(r^2)^3}\right.\nonumber\\
&& +\shalf\G(2+\l\e)\G(1-\e)\G(3-\l\e)
\frac{15\d_{(\m\n}r_{\r}r_{\s}r_{\t}r_{\p)}} {(r^2)^2} \nonumber \\
&& +\frac{1}{4}\G(1+\l\e)\G(2-\e)\G(2-\l\e)
\frac{45\d_{(\m\n}\d_{\r\s}r_{\t}r_{\p)}} {r^2} \nonumber \\
&& \left.+\frac{1}{8}\G(\l\e)\G(3-\e)\G(1-\l\e)
15\d_{(\m\n}\d_{\r\s}\d_{\t\p)}\right]
\label{b10}
\ena
The infrared subtractions in (\ref{b1})-(\ref{b10}) 
can then be performed using (\ref{a1})-(\ref{a4}).

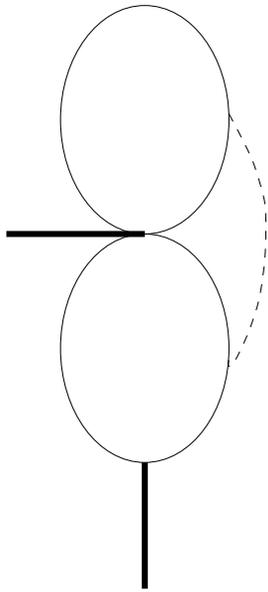
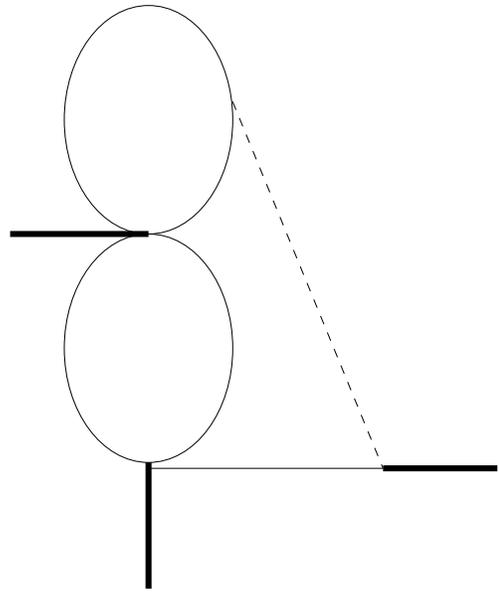
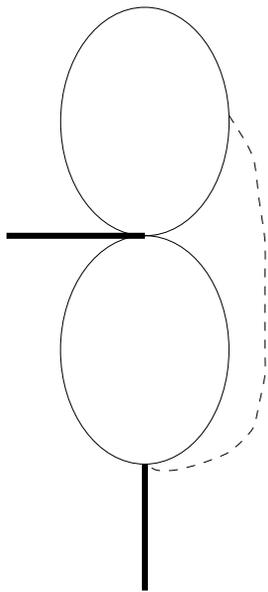
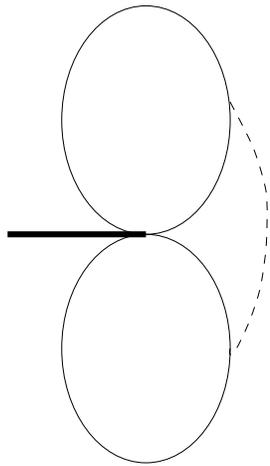
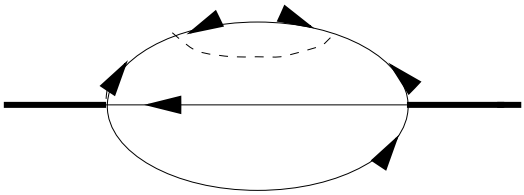
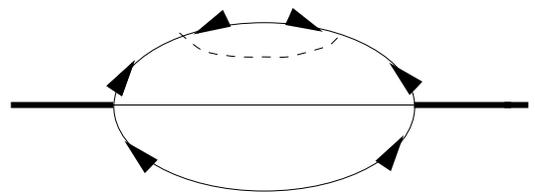
\end{document}